\documentclass[twocolumn,showpacs,preprintnumbers,amsmath,amssymb]{revtex4}

\usepackage{graphicx}
\usepackage{dcolumn}
\usepackage{bm}
\usepackage{subfigure}

\def\be {\begin{eqnarray} }

\def\ee {\end{eqnarray} }

\begin{document}


\title{Dynamics of interacting dark energy}

\author{Gabriela Caldera-Cabral}
\email{Gaby.Calderacabral@port.ac.uk}
\author{Roy Maartens}
\email{Roy.Maartens@port.ac.uk}
\affiliation{Insitute of Cosmology
\& Gravitation, University of Portsmouth, Portsmouth PO1 2EG, UK}

\author{L. Arturo Ure\~na-L\'opez}%
\email{lurena@fisica.ugto.mx}
\affiliation{Departamento de
F\'isica, DCI, Campus Le\'on, Universidad de
  Guanajuato, C.P. 37150, Le\'on, Guanajuato, M\'exico}%

\date{\today}

\begin{abstract}

Dark energy and dark matter are only indirectly measured via their
gravitational effects. It is possible that there is an exchange of
energy within the dark sector, and this offers an interesting
alternative approach to the coincidence problem. We consider two
broad classes of interacting models where the energy exchange is a
linear combination of the dark sector densities. The first class
has been previously investigated, but we define new variables and
find a new exact solution, which allows for a more direct,
transparent and comprehensive analysis. The second class has not
been investigated in general form before. We give general
conditions on the parameters in both classes to avoid unphysical
behavior (such as negative energy densities).

\end{abstract}

\pacs{98.80.-k 95.35.+d 95.36.+x 98.80.Jk}

\maketitle

\section{Introduction \label{sec:introduction}}

Cosmological observations point to the existence of non-baryonic
cold matter and of a late-time acceleration of the universe (see,
e.g.~\cite{Dunkley:2008ie,Tegmark:2006az,Percival:2007yw}). If gravity is modelled on
cosmological scales by general relativity, and if we assume that
the universe is homogeneous and isotropic on these scales, then
the late-time acceleration is sourced by a dark energy component,
and the universe is dominated by the ``dark sector". The quest to
uncover the true nature of dark matter (DM) and dark energy (DE)
is one of the most pressing topics of modern cosmology.

One of the fundamental puzzles within this quest is the
``coincidence problem": how is it that we seem to live in a time
when the densities of DM and DE are of the same order of
magnitude, given that they evolve very differently with redshift?
An interesting proposal is that interaction between the dark
fields could perhaps alleviate the coincidence problem. Various
interaction models have been put forward and studied (see,
e.g.~\cite{Wetterich:1994bg,Amendola:1999qq,Billyard:2000bh,
Zimdahl:2001ar,Dalal:2001dt,Farrar:2003uw,Chimento:2003iea,Cai:2004dk,Majerotto:2004ji,Nunes:2004wn,Olivares:2005tb,Sadjadi:2006qp,Guo:2007zk,GarciaCompean:2007vh,Setare:2007we,Boehmer:2008av,Quartin:2008px,Quercellini:2008vh,Valiviita:2008iv,He:2008si,Chen:2008ft}).

For a flat FRW universe, the background dynamics after
recombination are governed by the equations of energy balance and
the Raychaudhuri field equation:
\begin{subequations}
\label{eq:motion}
\begin{eqnarray}
  \dot{\rho}_b &=& - 3H \rho_b \, , \label{eq:motiond} \\
  \dot{\rho}_c &=& - 3H \rho_c + Q \, , \label{eq:motiona} \\
  \dot{\rho}_x &=& - 3(1+w_x) H \rho_x - Q \, ,
  \label{eq:motionb} \\
  \dot{H} &=& - 4\pi G \left[ \rho_b + \rho_c + (1+w_x) \rho_x
    \right] , \label{eq:motionc}
\end{eqnarray}
\end{subequations}
where $H = \dot{a}/a$ is the Hubble parameter, $\rho_c$ is the
cold DM density, $\rho_b$ is the baryonic density, $\rho_x$ is the
density of DE and $w_x <0$ is its constant equation of state. The
baryons only interact gravitationally with the dark sector, and
$Q$ is the rate of energy transfer in the dark sector. The
Friedmann constraint equation is
\begin{equation}
H^2 = \frac{8\pi G}{3} \left( \rho_b + \rho_c + \rho_x \right) \,.
\label{eq:friedmann}
\end{equation}
Note that the field equations~(\ref{eq:motionc}) and
(\ref{eq:friedmann}) are independent of $Q$, because of total
energy conservation. A positive $Q > 0$ represents transfer of
energy from DE to DM; a negative $Q < 0$ represents transfer of
energy from DM to DE.

In this paper, we consider interactions that are linear
combinations of the dark sector densities:
 \be
Q= A_c\rho_c + A_x \rho_x\,. \label{Q}
 \ee
Here the rate factors $A_I$ are either proportional to $H$ or
constants, leading to two classes of interaction model:
 \be
\mbox{ Model I}~ && A_I= 3\alpha_I H\,, \\
\mbox{ Model II} && A_I= 3\Gamma_I \,,
 \ee
where $\alpha_I$ are dimensionless constants and $\Gamma_I$ are
constant transfer rates. Observations impose the general
constraint that the interaction should be sub-dominant today, so that
 \be
|\alpha_I| \ll 1\,, ~~~~ |\Gamma_I| \ll H_0\,. \label{con}
 \ee

Model I has been recently analysed by~\cite{Quartin:2008px} (and
earlier work considered the special cases
$\alpha_c=\alpha_x$~\cite{Chimento:2003iea} and
$\alpha_x=0$~\cite{Billyard:2000bh}). We use an alternative
approach, defining new variables to simplify the parameter space,
and finding the general exact solution. We are able to recover
previous results more directly and simply, and to provide some new
insights into the model.

The $H$ in the $Q$-term for Model~I is motivated purely by
mathematical simplicity. By contrast, the energy exchange in
Model~II is motivated by similar models that have been used in
reheating~\cite{Turner:1983he}, curvaton decay~\cite{Malik:2002jb}
and decay of DM to radiation~\cite{Cen:2000xv}. As far as we know,
Model~II for the dark sector interaction has has not been treated
in the general case before. The special case $\Gamma_x=0$ has been
analyzed by~\cite{Valiviita:2008iv} (and by~\cite{Boehmer:2008av}
in the case where DE is modelled by exponential quintessence).

A summary of the paper is as follows. Some properties of the
general case ($Q$ not specified) are presented in
Sec.~\ref{sec:math-backgr}. Model~I is studied in
Sec.~\ref{sec:model-i.-q}. In Sec.~\ref{sec:model-ii.-q}, we
analyze Model~II. Finally, we conclude in
Sec.~\ref{sec:conclusions}.

\section{The case of general $Q$ \label{sec:math-backgr}}

We define the dimensionless dynamical variables\cite{Copeland:1997et}
\begin{equation}
  x = \frac{8\pi G}{3H^2} \rho_x \, , \quad        y = \frac{8\pi
  G}{3H^2} \rho_c \, , \quad z = \frac{8\pi G}{3 H^3} Q \, .
  \label{eq:variables}
\end{equation}
Before proceeding further, we now show that we can ``hide" the
constant DE equation of state by defining a new interaction
variable $\tilde{z} := z/(-w_x)$; in this way, all our results
below will be independent of the value of $w_x$.

The dark sector balance equations read
\begin{subequations}
  \label{eq:dyna-sys}
  \begin{eqnarray}
    x^\prime &=& 3 x \left( 1 - x \right) - \tilde{z} \,
    , ~~\tilde{z}:= -\frac{z}{w_x} \, , \label{eq:dyna-sysa} \\
    y^\prime &=& -3 xy + \tilde{z} \, ,
     \label{eq:dyna-sysb}
\end{eqnarray}
\end{subequations}
where a prime denotes $d/d(-w_x N)$, with $N=\ln a$ (we choose
$a_0=1$). The baryonic density is determined by the Friedmann
constraint,
\begin{equation}
\frac{8\pi G}{3H^2} \rho_b =   1 - x - y \, . \label{eq:fried}
\end{equation}

We cannot solve the system of equations until $Q$ is specified,
but we can draw some conclusions for a general $Q$ or $\tilde{z}$.
The simplest cases are $\tilde{z}=\tilde{z}(x,y)$, when the
system~(\ref{eq:dyna-sys}) is closed and autonomous. Model~I is
such a case. The next simplest cases are those for which $z$ is
not determined algebraically by $x$ and $y$, but does satisfy an
equation of motion of the form $\tilde{z}^\prime=F(x,y,\tilde{z})$, so
that we have a 3-dimensional autonomous system. Model~II is an
example of this case.

The critical points $(x_*,y_*)$ of the dynamical
system~(\ref{eq:dyna-sys}) must comply with the conditions
\begin{subequations}
  \label{eq:critical}
  \begin{eqnarray}
    3x_* \left( 1 - x_* - y_* \right) = 0 \,
    , \label{eq:criticala} \\
    -3x_* \left( 1 - x_* + y_* \right) + 2 \tilde{z}_* = 0 \,
    , \label{eq:criticalb}
  \end{eqnarray}
\end{subequations}
where $\tilde{z}_*$ is the interaction variable $\tilde z$
evaluated at the critical points. Equation~(\ref{eq:criticala})
implies that either $x_* = 0$, which represents the usual matter
dominated point, or $x_* + y_* = 1$, which implies no contribution
from the baryonic component.

From Eq.~(\ref{eq:criticalb}), we see that the option $x_* = 0$
directly implies that $\tilde{z}_* = 0$. Thus pure matter
domination can only exist if the interaction term vanishes at the
corresponding point. On the other hand, the option $x_* + y_* = 1$
leads to
\begin{equation}
  3x_* (1- x_*) = \tilde{z}_* \, . \label{eq:general-critical}
\end{equation}
If the dynamical system is autonomous, the above equation depends
only on $x_*$ and gives the position of the critical point that is
compatible with a nonzero interaction term (and no baryonic
contribution). By contrast, if the system is not autonomous, we
require extra information before we can completely determine its
critical points.

Another quantity of interest is the dark energy-to-dark matter
(DE-DM) ratio, $R = \rho_x /\rho_c$. From Eqs.~(\ref{eq:dyna-sys})
we obtain the evolution equation
\begin{equation}
  \frac{R^\prime}{R} = 3 - \frac{(x+y)}{xy} \tilde{z} \,
  ,~~ R= \frac{\rho_x}{\rho_c}= \frac{x}{y} \,, \label{DEtoDM}
\end{equation}
which is equivalent to Eq.~(3)
in~\cite{delCampo:2008sr,delCampo:2008jx}. As we show below, this
equation can lead to exact solutions in some particular models. A key
point relates to the factor $(x+y)$ in Eq.~(\ref{DEtoDM}): its
substitution by the Friedmann constraint~(\ref{eq:fried}) in some
cases can hide the existence of exact solutions
(e.g.~\cite{Barrow:2006hia,Quartin:2008px,delCampo:2008sr,delCampo:2008jx}). The
standard non-interacting case, $\tilde{z}=0$, shows the expected
exponential result $R=R_0 e^{-3w_x N}$, where $R_0
=\Omega_{x0}/\Omega_{c0}\sim 3.4$ is the present value of the DE-DM
ratio.

If the current value of $R$ is close to an asymptotic value, $R_0
\simeq R_\infty := R (\infty)$, then the coincidence problem is
``solved". (Strictly, one has transferred the coincidence problem
to a problem of explaining the dark sector interaction.) In this
case, $R$ must be a slowly-evolving function as $N \to \infty$,
i.e. $R^\prime \simeq 0$ at late times, which requires
\begin{equation}
  3xy - \tilde{z} \left( x + y \right) \simeq 0 \,
  . \label{slow-ratio}
\end{equation}
As we shall discuss below, for some models this condition is met
only at some points of the phase-space.

\section{Model I: $Q =  3 H \left( \alpha_x \rho_x + \alpha_c \rho_c
  \right)$ \label{sec:model-i.-q}}

This model was intensively studied in~\cite{Quartin:2008px} (our
$\alpha_I$ correspond to their $\lambda_I$). We find the critical
points in a more direct way, and also find the general exact
solution. This allows us to recover many of their results more
directly and simply, and to provide some new insights into the
model.

\subsection{Critical points and their properties \label{sec:critical-points-i}}

The dimensionless interaction variable $\tilde{z}$ is
\begin{equation}
  \tilde{z} = 3 \left( \tilde{\alpha}_x x + \tilde{\alpha}_c y \right)
  \, ,~~ \tilde{\alpha}_I = - \frac{\alpha_I}{w_x}\,. \label{eq:coupling}
\end{equation}
Thus $\tilde{z}=\tilde{z}(x,y)$ and the phase-space is
2-dimensional and autonomous.

As discussed before, the first type of critical point in
Eqs.~(\ref{eq:critical}) is that for which $x_* = 0$ and
$\tilde{z}_* =0$. Such a critical point in the present model also
needs $y_* =0$, and then it corresponds to a baryon dominated
point. This point is not realistic and we will exclude it from our
analysis; for more details, see~\cite{Quartin:2008px}.

The critical points that are compatible with a non-zero
interaction term comply with the constraint $x_* + y_* = 1$ and
are solutions of Eq.~(\ref{eq:general-critical}) in the form
\begin{equation}
  x_* \left( 1 - x_* \right) = \tilde{\alpha}_x x_* + \tilde{\alpha}_c
  \left( 1 - x_* \right) \, . \label{simple-sys}
\end{equation}
The critical points and their linear stability are summarized in
Table~I; the results are in agreement with those
of~\cite{Quartin:2008px}. For convenience we have defined the
parameters
\begin{subequations}
\label{eq:C}
\begin{eqnarray}
C_1 &=& \frac{1}{2} \left(1 - \tilde{\alpha}_x - \tilde{\alpha}_c
\right) \,
, \label{eq:C1} \\
C_2 &=& \sqrt{C^2_1 - \tilde{\alpha}_x \tilde{\alpha}_c} \, .
\label{eq:C2}
\end{eqnarray}
\end{subequations}
Note that the existence condition for the critical points ensures
that $C_2$ is real.

\begin{table*}[htp]
  \begin{center}
    \begin{tabular}{|c|c|c|c|c|c|} \hline
      Point &$x_*$ &$ y_*$ & Existence & Eigenvalues & Stability \\
      \hline \hline 
      &&&&& Unstable if $x_A < 0$ and $C_2 + \alpha_x > 0$ \\ 
      A & $C_1 + \tilde{\alpha}_c -C_2$ & $1-x_A$ & $C^2_2 \geq 0 $ &
      $-3(C_1 + \tilde{\alpha}_c - C_2)$; & Saddle if $x_A > 0$ and
      $C_2 + \alpha_x > 0$ \\ 
      & & & & $ 6(C_2 + \tilde{\alpha}_x)$ & or: $x_A < 0$ and $C_2 +
      \alpha_x < 0$ \\
      &&&&& Stable if $x_A > 0$ and $C_2 + \alpha_x < 0$ \\ 
      \hline 
      &&&&& \\
      B & $C_1 + \tilde{\alpha}_c +C_2$ & $1-x_B$ & $C^2_2 > 0$ &
      $-3(C_1+\tilde{\alpha}_c+C_2)$; & Saddle if $x_B < 0$ \\
      &&&& $-6 C_2$ & Stable if $x_B > 0$ \\
      &&&&& \\
      \hline
    \end{tabular}
  \end{center}
  \caption[crit]{\label{crit1} Critical points of Model~I and their
    stability. $C_1,C_2$ are given in Eqs.~(\ref{eq:C}); notice that
    $C_2$ is positive by definition. The existence conditions
      are given here in general, but the discussion in the text
      suggests that both interaction parameters should be positive,
      see Eq.~(\ref{eq:positive}), so that both $x_A$ and $x_B$ are
      positive too. Thus, point A is of the saddle type whereas point
      B is stable.}
\end{table*}

For a physically viable model, one of the critical points should
correspond to a DM dominated universe, $x \to 0$ and $y \to 1$, at
early enough times; moreover, this critical point should be an
unstable point. The only candidate is point A, but $x_A\neq 0$.
For DM domination we need $|x_A| \ll 1$; to linear order in $x_*$,
we obtain from Eq.~(\ref{simple-sys}) that
\begin{equation}
  x_A \simeq \frac{\tilde{\alpha}_c}{1-\tilde{\alpha}_x} \, , \quad
  y_A \simeq \frac{2C_1}{1 - \tilde{\alpha}_x} \,
  . \label{eq:smallDE}
\end{equation}
At late times we must get a DE dominated universe, which should
correspond to the stable point B. For the latter,
Eq.~(\ref{simple-sys}) gives the estimate
\begin{equation}
  x_B \simeq \frac{2C_1}{1 - \tilde{\alpha}_c} \, ,
  \quad y_B \simeq \frac{\tilde{\alpha}_x}{1-\tilde{\alpha}_c} \,
  . \label{eq:smallDM}
\end{equation}

For certain parameter values the DE density is negative at early
times~\cite{Quartin:2008px} (see also~\cite{Valiviita:2008iv} in
the cases $\alpha_x=0$ and $\alpha_x=\alpha_c$). In a first
approximation we can use Eqs.~(\ref{eq:smallDE})
and~(\ref{eq:smallDM}) to determine the conditions under which
both DM and DE are non-negative and well behaved at all times.

At early times, we must impose the constraint
\begin{equation}
  0 \leq x_A \simeq \frac{\tilde{\alpha_c}}{1 - \tilde{\alpha}_x} < 1
  \, ,. \label{eq:smallRA}
\end{equation}
which is satisfied if
\begin{equation}
  \tilde{\alpha}_c \geq 0 \, , \quad 1 > \tilde{\alpha}_x +
  \tilde{\alpha}_c \, , \label{eq:positive}
\end{equation}
or
\begin{equation}
  \tilde{\alpha}_c \leq 0 \, , \quad 1 < \tilde{\alpha}_x +
  \tilde{\alpha}_c \, . \label{eq:negative}
\end{equation}
Likewise, at late times we have the constraint
\begin{equation}
  0 \leq y_B \simeq \frac{\tilde{\alpha}_x}{1 - \tilde{\alpha}_c} < 1
  \, , \label{eq:smallRB}
\end{equation}
and then the same conditions~(\ref{eq:positive})
and~(\ref{eq:negative}) apply, but now with the interaction
constants interchanged, $\tilde{\alpha}_c \leftrightarrow
\tilde{\alpha}_x$. It can be verified that the conditions
described above give the correct description for the different
cases depicted in Figs.~7, 8 and 9 in~\cite{Quartin:2008px}.

It follows that for a plausible scenario, the interaction
parameters $\tilde{\alpha}_c$ and $\tilde{\alpha}_x$ must be both
non-positive or both non-negative. A remarkable result arises now.
If the interaction parameters are to have the same sign, the only
permitted case is that of Eq.~(\ref{eq:positive}), as the case in
Eq.~(\ref{eq:negative}) cannot be consistently satisfied if the
interaction parameters are both negative.

Another constraint appears from the condition that the critical
points are such that $0 \leq x_*,y_* \leq 1$. From
Eqs.~(\ref{eq:smallDE}), we get $\tilde{\alpha}_x < 1$; likewise,
from Eqs.~(\ref{eq:smallDM}) we get $\tilde{\alpha}_c < 1$.
Finally, the critical points should be real, $C^2_2 \geq 0$, and
then the allowed values $(\tilde{\alpha}_c , \tilde{\alpha}_x)$
are those located \emph{below} the rotated parabola
\begin{equation}
  (\tilde{\alpha}_x - \tilde{\alpha}_c)^2 \geq 2 (\tilde{\alpha}_x +
  \tilde{\alpha}_c) - 1 \, . \label{eq:parabola}
\end{equation}

Therefore, we arrive at the overall conclusion that $0 \leq
\tilde{\alpha}_c \, , \tilde{\alpha}_x \leq 1$ if the DM and DE
contributions are to be positive and well behaved, $0< x, y < 1$,
at all times. This is in agreement with the results
of~\cite{Quartin:2008px} (see the critical points $A_2$ and $B_2$
in Table~I). We used a first order calculation based on
Eqs.~(\ref{eq:smallDE}) and~(\ref{eq:smallDM}); however, by
continuity, we expect the result to be true even in cases where DE
and DM contribute significantly at early and late times.

A more general statement exists regarding the positivity of the
dark components at early times. Note that we can write
Eq.~(\ref{simple-sys}) for the critical points in the form
\begin{equation}
  3w_x x_* y_* = -3 \left( \alpha_x x_* + \alpha_c y_* \right) = - z_*
  \, . \label{eq:simple-critical}
\end{equation}
If both DM and DE are to be positive for all times, then $z$ and
$w_x$ should have opposite signs. This explains why the DE
component becomes negative at early times in certain models: it is
because the interaction variable $z$ can become negative at DM
domination. In our case, non-negative interaction constants ensure
that both DM and DE are positive at all times (see also Figs.~7
and 9 in~\cite{Quartin:2008px} for some other examples.)

\begin{figure}[!ht]
  \includegraphics[width=8.7cm,height=8cm]{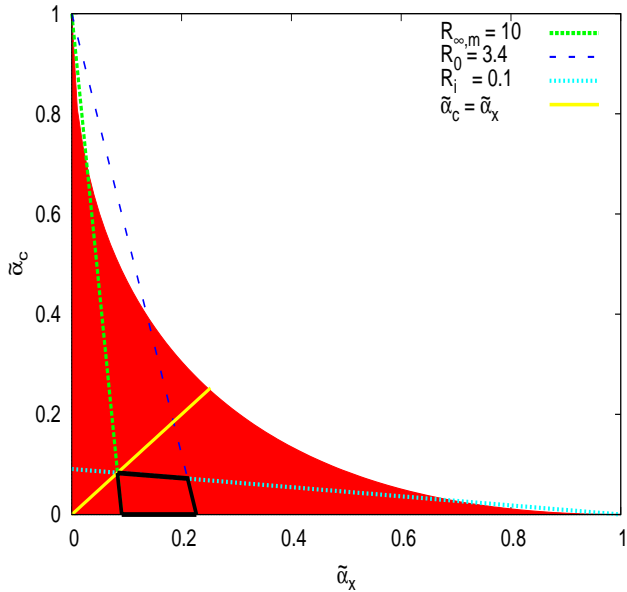}
  \caption{\label{fig:parameters}
    The red shaded region contains the allowed non-negative values of
    $\tilde{\alpha}_c$ and $\tilde{\alpha}_x$ satisfying the reality
    constraint~(\ref{eq:parabola}). The cyan (dotted) line represents
    the equality in Eq.~(\ref{eq:const1}) for $R_i=0.1$, the blue
    (dashed) line represents the equality in Eq.~(\ref{eq:const2}) for
    $R_0=3.4$, and the green (dashed-dotted) line is
    Eq.~(\ref{eq:const3}) with $R_{\infty,m}=10$. The yellow (solid)
    line corresponds to the particular case $\alpha_c =
    \alpha_x$~\cite{Chimento:2003iea}. The only region that provides
    reasonable values of the DE-DM ratio at both early and late times
    is the one surrounded by the thick black (solid) lines. The value
    $R_i=0.1$ was chosen for presentation purposes; the allowed region
    would become much smaller for a more realistic value $R_i \ll
    0.1$.}
\end{figure}

In Fig.~\ref{fig:parameters}, we show the allowed values of the
interaction parameters (represented by the red shaded region)
according to the discussion above.

\subsection{DE-DM ratio: exact solution}

With the interaction term~(\ref{eq:coupling}), Eq.~(\ref{DEtoDM})
becomes
\begin{equation}
R^\prime = -3 \left[ \tilde{\alpha}_x R^2 - \left( 1 -
  \tilde{\alpha}_x - \tilde{\alpha}_c \right) R + \tilde{\alpha}_c
  \right] \, . \label{eq:diffR}
\end{equation}
This equation provides the known solutions of simpler cases, e.g.
for the case $\alpha_x=0$, we have
 \be
R=\left(\frac{\Omega_{x0}}{\Omega_{c0}} +
\frac{\alpha_c}{w_x+\alpha_c} \right) e^{-3(w_x+\alpha_c)N} -
\frac{\alpha_c}{w_x+\alpha_c} \, ,
 \ee
which directly recovers Eq.~(7) of~\cite{Guo:2007zk}. Here we
present the new solution in the general case $\alpha_x \neq0$:
\begin{equation}
R(N) = \frac{C_1}{\tilde{\alpha}_x} - \frac{C_2}{\tilde{\alpha}_x}
\tanh \left[ 3 w_x C_2 \left(N - N_1
  \right) \right] \, , \label{eq:simple-ratio}
\end{equation}
where $C_1$ and $C_2$ are given by Eqs.~(\ref{eq:C}), and the
integration constant is \be N_1 = - \frac{1}{3w_x C_2} \tanh^{-1}
\left[ \frac{1}{C_2}
  \left(\tilde{\alpha}_x \, \frac{\Omega_{x0}}{\Omega_{c0}} -
  C_1\right) \right] \, . \label{eq:observables0}
\ee Using Eq.~(\ref{eq:simple-ratio}) we can integrate the DM
energy balance equation to obtain the DM density as a function of
$N$:
 \be
\rho_c = \rho_{c0}e^{(-3-\alpha_c-C_1)N}\left[ \frac{\cosh  3C_2
    (N-N_1) }{\cosh  3C_2N_1 } \right]^{1/3}
, \label{eq:rho-c}
 \ee
and then the DE density follows directly as
$\rho_x(N)=R(N)\rho_c(N)$.

The asymptotic values of the DE-DM ratio from
Eq.~(\ref{eq:simple-ratio}) are directly related to the values
inferred from the critical points in Table~\ref{crit1},
\begin{subequations}
\begin{eqnarray}
  R_{-\infty} &:=& R(-\infty) = \frac{C_1 - C_2}{\tilde{\alpha}_x}
  = \frac{x_A}{y_A} \, , \\
  R_\infty &:=& R(\infty) = \frac{C_1 + C_2}{\tilde{\alpha}_x} =
  \frac{x_B}{y_B} \, . \label{eq:observables}
\end{eqnarray}
\end{subequations}
It follows that the smallest value of the DE-DM ratio
$R_{-\infty}$ is determined by the critical point A; likewise, its
largest value $R_\infty$ is determined by the critical point B.
The correspondence between the asymptotic values of $R$ and the
critical points A and B is not surprising after all, because the
roots of $R^\prime =0$ in Eq.~(\ref{eq:diffR}) are actually the
ratios inferred from the critical points.

There is an interesting point concerning the initial values of the
DE-DM ratio. For given values of the free parameters of the model
($w_x$, $\alpha_x$, and $\alpha_c$), the initial value of the
DE-DM ratio, $R_i$, should be such that
\begin{equation}
  R_i > R_{-\infty} \, , \label{eq:Rpast}
\end{equation}
otherwise the evolution of the DE-DM ratio is not described by
Eq.~(\ref{eq:simple-ratio}). A similar constraint exists for the
late time value $R_\infty$:
\begin{equation}
  R_\infty > R_0 \, , \label{eq:Rfuture}
\end{equation}
to ensure that the value $R_0$ is included in the range of values
allowed by the exact solution~(\ref{eq:simple-ratio}).

Further constraints on the values of the interaction parameters
can be obtained from Eqs.~(\ref{eq:Rpast}) and (\ref{eq:Rfuture}).
According to Eq.~(\ref{eq:smallDE}), if the DE contribution is to
be small at early times, we can write
\begin{equation}
  R_{-\infty} = \frac{x_A}{y_A} \simeq \frac{\tilde{\alpha}_c}{2C_1}
  \, , \label{eq:early-ratio}
\end{equation}
and then, for a given value of $R_i$, Eq.~(\ref{eq:Rpast}) becomes
the constraint
\begin{equation}
  \tilde{\alpha}_c < \frac{R_i}{1+R_i} ( 1- \tilde{\alpha}_x ) \,
  . \label{eq:const1}
\end{equation}

By using Eq.~(\ref{eq:smallDM}), i.e. under the assumption that
the DM contribution is small at late times, we find the companion
expression of Eq.~(\ref{eq:early-ratio}), which is
\begin{equation}
  R_\infty = \frac{x_B}{y_B} \simeq \frac{2C_1}{\tilde{\alpha}_x} \,
  , \label{eq:late-ratio}
\end{equation}
and then Eq.~(\ref{eq:Rfuture}) becomes the constraint
\begin{equation}
    \tilde{\alpha}_c < 1 - (1+R_0) \, \tilde{\alpha}_x \, . \label{eq:const2}
\end{equation}

If we want an upper limit on $R_\infty$, i.e. $R_\infty <
R_{\infty,m}$, where $R_{\infty,m}$ is some maximum value, then a
new constraint arises from Eq.~(\ref{eq:late-ratio}),
\begin{equation}
  \tilde{\alpha}_c > 1 - (1+R_{\infty,m}) \, \tilde{\alpha}_x \,
  . \label{eq:const3}
\end{equation}

The inclusion of the constraint equations~(\ref{eq:const1}),
(\ref{eq:const2}), and (\ref{eq:const3}) in
Fig.~\ref{fig:parameters} indicates that only a small region of
the parameter space may be compatible with observations. The
examples in Fig.~\ref{fig:parameters} correspond to $R_i =0.1$,
$R_0=3.4$ and $R_{\infty,m}=10$.

It can be verified that the above results are in agreement with
the results presented
in~\cite{Quartin:2008px,Chimento:2003iea,Guo:2007zk,
Valiviita:2008iv,Olivares:2007rt,Barrow:2006hia}. In particular,
the diverse cases presented in those papers are explained in a
unified way by the exact solution~(\ref{eq:simple-ratio}), and our
approach provides very simple expressions for the analysis of the
parameter space. For example, it readily explains the troublesome
features encountered in Figs.~7, 8 and 9 of~\cite{Quartin:2008px}.

\begin{figure}[htp]
  \includegraphics[width=8.7cm,height=8cm]{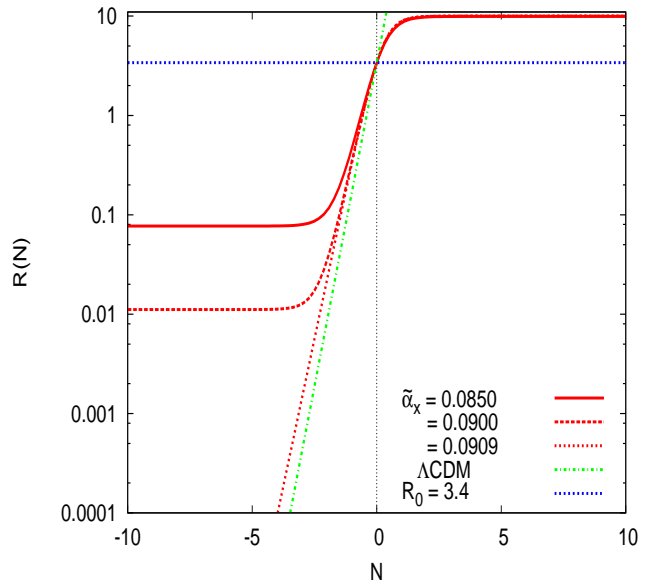}
  \caption{\label{fig:dedmratio}
    Evolution of the DE-DM ratio $R(N)$ according to
    Eqs.~(\ref{eq:simple-ratio}) and (\ref{eq:C}). The chosen values
    of the various parameters are $R_0 =3.4$, $R_{\infty}=10$,
    whereas the value of $\tilde{\alpha}_c$ was determined from
    Eq.~(\ref{eq:late-ratio}) for the given values of
    $\tilde{\alpha}_x$. For comparison, the green (dashed-dotted) line
    represents the standard $\Lambda$CDM case. The dotted line
    corresponds to exactly $\tilde{\alpha}_c=0$, for which case
    $R_{-\infty}=0$ and the earlier evolution is very similar to that
    of standard $\Lambda$CDM.}
\end{figure}

\begin{figure}[htp]
  \includegraphics[width=8.7cm,height=8cm]{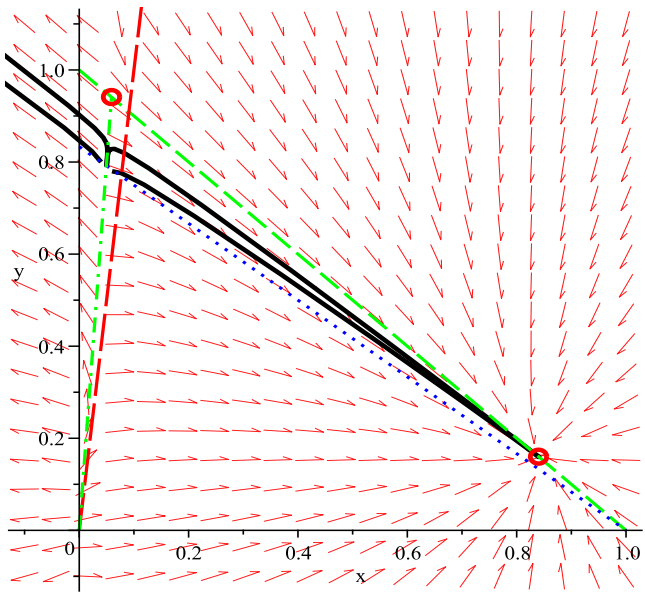}
  \caption{\label{fig:phasespace}
  The phase space of Model~I in the case $\tilde{\alpha}_x =0.15$, and
  $\tilde{\alpha}_c =0.05$. The (red) circles are the critical points
  A and B (see Table~\ref{crit1}), and the green dashed line is the
  (heteroclinic) constraint $x+y=1$ that connects them. The green
  dot-dashed line that connects the saddle point A to the unstable
  critical point at the origin is a good approximation to the
  heteroclinic trajectory between the two points. The red long-dashed
  line represents the ratio $x/y = 0.1$. The blue dotted line is the
  approximate Friedmann constraint~(\ref{eq:fried-aprox}). The black
  (solid) lines are numerical solutions of the equations of motion for
  different initial conditions. The trajectories with initial
  conditions on the right of the heteroclinic line end up at point B;
  if the initial conditions are on the left, the DE component becomes
  negative and the DM one grows without bound.}
\end{figure}

In Fig.~\ref{fig:dedmratio} we show examples of the evolution of
the DE-DM ratio for fixed values $R_{\infty}$, $R_0$, and $w_x$.
The curves correspond to different values of $\tilde{\alpha}_x$,
and the values of $\tilde{\alpha}_c$ were determined from
Eq.~(\ref{eq:late-ratio}). Finally, we show in
Fig.~\ref{fig:phasespace} a typical example of a phase space with
parameters in the allowed region of Fig.~\ref{fig:parameters}.

We are assuming that the initial conditions are set at the onset
of matter domination, and that the DM to baryonic matter ratio is
the same as in the standard case,
\begin{equation}
  \left. \frac{\rho_b}{\rho_c} \right|_{i} \simeq
  \frac{\Omega_{b,0}}{\Omega_{m,0}} \simeq \frac{1}{5} \, .
\end{equation}
Note that the above is just an approximation, since the DM
component in the interacting case does not evolve exactly as
$a^{-3}$, see Eq.~(\ref{eq:rho-c}). Then the initial conditions of
the DM and DE contributions, with the help of the Friedmann
constraint~(\ref{eq:fried}), are determined from
\begin{equation}
  x_i + \frac{6}{5} y_i \simeq 1 \, . \label{eq:fried-aprox}
\end{equation}

Some examples of numerical solutions of the equations of motion
are shown in Fig.~\ref{fig:phasespace} for different initial
conditions. All of them represent valid solutions that start at a
matter dominated epoch and end in a final state with finite DE-DM
ratio. This final state is uniquely determined by the values of
the free parameters of the model.

The phase space portrait reveals some other aspects of the
interacting model under consideration, apart from the critical
points and their stability studied above. It shows the line
$R_{-\infty} = x/y$, which approximates very well the heteroclinic
curve that connects the (unstable) critical point at the origin
(actually, this is the baryonic dominated critical point discussed
in Sec.~\ref{sec:critical-points-i}) with the (saddle) point A. It is
then apparent that any curve with initial DE-DM ratio $R_i <
R_{-\infty}$ leads to trajectories in which the DE component is
negative and the DM grows without bound.

Therefore, the constraint equation~(\ref{eq:Rpast}) is not only
necessary for the trajectory to be described in terms of the DE-DM
ratio in Eq.~(\ref{eq:simple-ratio}), but it is also needed to
have a reasonable evolution of the universe after the onset of
matter domination.

The initial value $R_i$ is related to the e-fold number $N_{i}$ by
\begin{equation}
    R_i = - \frac{C_1}{\alpha_x} + \frac{C_2}{ \alpha_x} \tanh \left[
    3C_2 \left(N_{i} - N_1 \right) \right] \, ,
\end{equation}
where $N_1$ is determined from Eq.~(\ref{eq:observables0}), so
that there is a one-to-one correspondence between $R_i$ and
$N_{i}$. Thus different initial conditions result in different
times for the appearance of a matter dominated epoch. Different
initial conditions can have dramatic effects on the past history
of the universe, even if the final state can be arranged to be the
same in all cases.

\subsection{Special cases}

Our results apply for some special cases already present in the
literature. For instance, if $\tilde{\alpha}_x =
\tilde{\alpha}_c$\cite{Chimento:2003iea}, then Eqs.~(\ref{eq:C})
suggest that $\tilde{\alpha}_c \leq 1/4$ for the critical points
to be real (see also Eq.~(\ref{eq:parabola})). Moreover,
Eq.~(\ref{eq:smallDE}) implies that $\tilde{\alpha}_c \ll 1$ if
proper matter domination is to appear and then $R_A \ll 1$. Since
Eqs.~(\ref{eq:smallRA}) and~(\ref{eq:smallRB}) together imply that
$R_B = 1/R_A$, this model has difficulty to achieve an
asymptotically constant DE-DM ratio relevant to the coincidence
problem.

Another example is $\tilde{\alpha}_x=0$~\cite{Guo:2007zk}, for
which the critical points, from Eq.~(\ref{simple-sys}), are $x_A=
\tilde{\alpha}_c$ and $x_B=1$. However, Fig.~\ref{fig:parameters}
indicates that any reasonable model necessarily needs
$\tilde{\alpha}_x \neq 0$ to alleviate the coincidence problem.

For completeness, we can also have the case $\tilde{\alpha}_c=0$.
Then there is proper matter domination for any value $0 \leq
\tilde{\alpha}_x \leq 1$, because $x_A=0$, and the case is free
from the problems related to a finite value of $R_{-\infty}$, see
Figs.~\ref{fig:parameters}, \ref{fig:dedmratio},
and~\ref{fig:phasespace}. The coincidence problem can be addressed
if the only non-zero interaction parameter is given an
appropriately small value, as $R_B \simeq 1/\tilde{\alpha}_x$.

\section{Model II. $Q = 3 \left( \Gamma_x \rho_x + \Gamma_c \rho_c
  \right)$ \label{sec:model-ii.-q}}

In the simplest model of the reheating process after inflation in
the early universe, one assumes that the oscillating inflaton
field $\phi$ behaves like a matter fluid that decays into
relativistic particles; the decay is parametrized by a constant
decay width $\Gamma_\phi$~\cite{Turner:1983he}. In our notation,
$Q = \Gamma_\phi \rho_\phi$. Motivated by this, and by similar
models for curvaton decay~\cite{Malik:2002jb} and for decay of DM
to radiation~\cite{Cen:2000xv}, we arrive at Model~II: $Q =3(
\Gamma_x \rho_x + \Gamma_c \rho_c)$, where the $\Gamma_I$ are
constant decay widths. Unlike Model~I, the $Q$ here is not constructed
a priori for mathematical simplicity, and the dynamics are
considerably more complicated as a result.

The variable $z$ becomes
\begin{equation}
z = 3 \left( \frac{\Gamma_x}{H} x + \frac{\Gamma_c}{H} y \right)
\, , \label{eq:model2-z}
\end{equation}
which typically grows in an expanding universe, and diverges in
the limit $H \to 0$. Because of this, it is convenient to define
the new variable~\cite{Boehmer:2008av},
\begin{equation}
  u := \frac{H_0}{H + H_0} \, , \label{eq:u-variable}
\end{equation}
which allows us to compactify the evolution of $z$. Here, $H_0$
denotes the current value of the Hubble parameter. Early times
correspond to $u \to 0$, and late times to $u \to \rm{const}$ (if
$\dot H<0$, then $u \to 1$). As in Model I, we redefine the
interaction variables as
\begin{equation}
\tilde{\alpha}_I (u) := \frac{u}{1-u} \frac{\gamma_I}{(-w_x)} \, ,
\quad \gamma_I:= \frac{\Gamma_I}{H_0} \, . \label{gamma}
\end{equation}
Notice that $u_0=1/2$ and then
$\tilde{\alpha}_{I0}=\gamma_I/(-w_x)$. Thus
\begin{equation}
\tilde{z} = \frac{3u}{1-u} \frac{( \gamma_x x + \gamma_c
y)}{(-w_x)}  = 3 \left[ \tilde{\alpha}_x(u) \, x +
\tilde{\alpha}_c (u) \, y \right] \, ,
\end{equation}
which is a time-dependent version of the Model~I
expression~(\ref{eq:coupling}). The equations of motion become
\begin{subequations}
  \label{eq:dyna-sysmod2}
  \begin{eqnarray}
    x^\prime &=& 3 x \left( 1 - x \right) - 3 \left[
      \tilde{\alpha}_x(u) \, x + \tilde{\alpha}_c (u) \, y \right] \,
    , \label{eq:dyna-sysamod2} \\
    y^\prime &=& -3 xy + 3 \left[ \tilde{\alpha}_x(u) \, x +
      \tilde{\alpha}_c (u) \, y \right] \, , \label{eq:dyna-sysbmod2}
    \\
    u^\prime &=& -\frac{3}{2w_x} ( 1 + w_x x ) u (1-u) \,
    , \label{eq:dyna-syscmod2}
 \end{eqnarray}
\end{subequations}
where a prime again denotes $d/d(-w_x N)$. Note that the DE
equation of state appears explicitly in
Eq.~(\ref{eq:dyna-syscmod2}) because the value of $w_x$ is
necessary to know the time evolution of the cosmological model.

\subsection{Critical points \label{sec:critical-points-}}

The system~(\ref{eq:dyna-sysmod2}) is autonomous, and, to begin
with, Eq.~(\ref{eq:dyna-syscmod2}) admits the critical values
$u_*=0$ and $u_*=1$.

If $u_*=0$, then Eqs.~(\ref{eq:dyna-sysamod2})
and~(\ref{eq:dyna-sysbmod2}) suggest the critical values:
(a)~$x_*=0$, with $y_*$ to be determined from the Friedmann
constraint~(\ref{eq:fried}); and (b)~$x_*=1$ and $y_*=0$. These
are the expected critical points from the general discussion in
Sec.~\ref{sec:math-backgr}, see Eqs.~(\ref{eq:critical}). The
stability of these points can be established by standard methods,
from which we find that (a) is unstable, while (b) is of saddle
type for $-1 < w_x < 0$ and stable for $w_x < -1$.

\begin{table*}[ht!]
\begin{center}
\begin{tabular}{|c|c|c|c|c|c|c|} \hline
  Point & $x_*$ & $y_*$ & $u_*$  & Existence & Eigenvalues & Stability
  \\
  \hline \hline 
  &&&&&& \\ 
  A & $0$ & $y_*$ & $0$ & all $ \gamma_c \, , \, \gamma_x$ &
  $0;\,\, 3;\,\, -3/(2w_x)$ & Unstable if $w_x < 0$ \\ 
  &&&&&& \\ 
  \hline 
  &&&&&& \\
  B & $1$ & $0$ & $0$ & all $ \gamma_c \, , \, \gamma_x$ &
  $-3(1+w_x)/(2w_x);$ & Saddle if $-1 < w_x < 0$ \\ 
  &&&&& $-3;\,\, -3$ & Stable if $w_x < -1$ \\
  &&&&&&\\
  \hline 
  &&&&&&\\
  C & $-\gamma_c / (\gamma_x - \gamma_c )$ & $\gamma_x / (\gamma_x -
  \gamma_c )$ & $1$ & $\gamma_c \neq \gamma_x$ &
  $\frac{3}{2w_x} \frac{\gamma_x - \gamma_c(1+w_x)}{\gamma_x -
    \gamma_c};$ &
  Stable if $-w_x x_C < 1$, and $x_C > 0$, and $\gamma_x > \gamma_c$
  \\
  &&&&& $3\gamma_c/(\gamma_x - \gamma_c);$ & Unstable otherwise \\
  &&&&& $-\infty \cdot \textrm{sgn}(\gamma_x - \gamma_c)$ & \\
  &&&&&& \\
  \hline
\end{tabular}
\end{center}
\caption[crit]{\label{crit2} Critical points of Model II. The early
  universe is $u\to 0$, the late universe is $u \rightarrow 1$, and
  $\gamma_I=\Gamma_I/H_0$. The $\infty$-eigenvalue for point C appears
  due to the limit $u \rightarrow 1$. As discussed in the text, see
  for instance Eq.~(\ref{eq:GammaC}), $\gamma_x > 0$ and $\gamma_c <0$
are required to have non-negative dark components at late times; those
same conditions directly imply that point C is stable.}
\end{table*}

On the other hand, for $u_*=1$ the only possibility, again from
the general discussion in Sec.~\ref{sec:math-backgr}, is $x_* +
y_* =1$, and then the critical value $x_*$ is determined from
Eq.~(\ref{simple-sys}) but with variable $\alpha_I$.
Equation~(\ref{simple-sys}) has solutions
\begin{equation}
  x^{\pm}(u) = C_1(u) \mp C_2(u) + \tilde{\alpha}_c(u) \,
  , \label{eq:quadraticIII}
\end{equation}
where $C_I(u)$ are defined as in Eq.~(\ref{eq:C}), with
$\tilde{\alpha}_I \to \tilde{\alpha}_I(u)$.

For $u \to 1$ (that is, $\tilde{\alpha}_I \to \infty$), the
asymptotic approximate solutions are
\begin{eqnarray}
  x^{\pm} &\simeq & \frac{1}{2} + \frac{u}{2w_x (1-u)} \left[ \left(
    \gamma_x - \gamma_c \right) \pm \left| \gamma_x - \gamma_c
    \right| \right]\nonumber \\
  &&~ \pm \frac{1}{2} \left( \frac{\left| \gamma_x -
    \gamma_c \right|}{\gamma_x - \gamma_c} \right) \left(
  \frac{\gamma_x + \gamma_c}{\gamma_x - \gamma_c} \right) \,
  . \label{eq:III-asymptotic}
\end{eqnarray}
The asymptotic values in the above equation depend on the values
of the interaction parameters. In the case $(\gamma_x - \gamma_c)
> 0$, we find
\begin{equation}
x^\pm\! \to\!  x^{\pm}_*= \left\{\!
  \begin{array}{c}
    -\infty \\
    \gamma_c /( \gamma_c - \gamma_x )
  \end{array} \right.\, , \quad y^{\pm}_* = 1- x^{\pm}_*  \,
  , \label{eq:III-asymptotic1a}
\end{equation}
and for $(\gamma_x - \gamma_c) < 0$, we find
\begin{equation}
x^\pm\! \to\!  x^{\pm}_*= \left\{\!
  \begin{array}{c}
    \gamma_c /( \gamma_c - \gamma_x ) \\
    \infty
  \end{array} \right. \, , \quad y^{\pm}_* = 1- x^{\pm}_*  \,
  . \label{eq:III-asymptotic1b}
\end{equation}
The critical points $(x^+_*,y^+_*,1)$ are always unstable, whereas
$(x^-_*,y^-_*,1)$ are always stable. The finite critical points of
the system~(\ref{eq:dyna-sysmod2}) and their stability are
summarized in Table~\ref{crit2}.

It should be stressed that Model~II is a time-dependent
generalization of Model~I, and then there should be good agreement
in the formulas of the two cases. For instance, a quick comparison
between Eqs.~(\ref{eq:smallDE}) and~(\ref{eq:smallDM}) and
Eqs.~(\ref{eq:III-asymptotic1a}) and~(\ref{eq:III-asymptotic1b}),
quickly shows this is the case.

Point A represents matter (DM and baryons) domination at early
times, with no contribution from DE (unlike in Model~I). Point C
is a late-time attractor only in the case $(\gamma_x - \gamma_c) >
0$, and then we also require $\gamma_c < 0$ if the energy density
of the dark fluids is to be positive at the critical point.
Furthermore, C is a scaling point, since the asymptotic DE-to-DM
ratio is
\begin{equation}
  R = \frac{x}{y} \to - \frac{\gamma_c}{\gamma_x} = -
  \frac{\Gamma_c}{\Gamma_x} \, , \label{eq:model2-Rfinal}
\end{equation}
which only depends on the ratio of the decay widths. This
asymptotic ratio also shows that the interaction rates $\Gamma_I$
should have opposite signs if both dark densities are to remain
positive at late times. Thus
 \be
\Gamma_x \geq 0 ~~\mbox{and}~~\Gamma_c \leq 0 \,, \label{eq:GammaC}
 \ee
are necessary conditions to have a finite and positive late-time
attractor in the model.

Apart from finding the critical points, we need to know the
behavior of the system at early times, since Model~II can also
exhibit a negative DE component, as shown
in~\cite{Valiviita:2008iv} in the case $\Gamma_x=0$.

In principle, we would need to scan exhaustively the 3-dimensional
phase space. A short-cut we will take is to find the points
$\hat{x}$ at which $x^\prime(\hat{x})=0$, see
Eq.~(\ref{eq:dyna-sysamod2}), for fixed values of the variables
$y$ and $u$. The result is
\begin{equation}
  \hat{x}^\pm(u) = \frac{1}{2} \left[ (1-\tilde{\alpha}_x) \pm
    \sqrt{(1-\tilde{\alpha}_x)^2 - 4 \tilde{\alpha}_c y} \right] \, .
\end{equation}
In the early universe, $u \to 0$, one possibility is
$\hat{x}^+=1$, but the interesting solution is
\begin{equation}
  \hat{x}^- \simeq \tilde{\alpha}_c \, y \, .
\end{equation}
Clearly, the value of $\hat{x}^-$ above marks the point at which
the $x$-component of the phase space velocity $x^\prime$ changes
sign.

If the evolution of the cosmological system departs from an early
unstable point corresponding to non-negative dark components, then
we must impose the condition $\hat{x}^- \geq 0$, otherwise the DE
variable will approach point A from below, $x \to 0^-$. This
generalizes the $\Gamma_x=0$ result of~\cite{Valiviita:2008iv} to
any value of $\Gamma_x$ and $w_x$ (see also the discussion below).
As a consequence, a non-negative DM interaction $\Gamma_c \geq 0$
is necessary to ensure a positive DE density at early times.

It is now clear that we cannot, in general, find a version of
Model~II in which negative values of the dark energy densities
are consistently avoided at both early and late times. In this sense,
Model~II can only be used if we restrict it to apply either (a)~from
the beginning only up to the present time, or (b)~from some finite
time (e.g. recombination) onwards.

\subsection{Special cases \label{sec:special-cases-}}

The first special case is $\Gamma_x = \Gamma_c$. Following the
discussion in the previous section, it is necessary to have
positive interaction rates to have a proper early matter era. For
the late Universe, the critical points are
\begin{equation}
  x^\pm_* = \frac{1}{2} \pm \frac{1}{2} \sqrt{1 -
    4\tilde{\alpha}_c(u)} \, ,
\end{equation}
and then a negative value of $\Gamma_c$ is required for the
critical points to be real. But even in this case the values of
$x^\pm_*$ are not finite in the limit $u \to 1$; hence, the simple
case $\Gamma_x = \Gamma_c$ cannot provide a realistic model.

In our numerical experiments we have found that the case
$(\gamma_x - \gamma_c) < 0$ has interesting properties. Firstly, a
problematic early evolution can be avoided if we choose $\gamma_c
\geq 0$; note that $\gamma_x$ may positive or negative, as long as
$\gamma_x < \gamma_c$. As $u \to 1$, the point C is unstable and there
are two possibilities for its late time evolution, see
Eq.~(\ref{eq:III-asymptotic1b}).

The first one arises if at some time the DE variable $x$ is to the
\emph{right} of point C, so that the system can freely follow the
late-time attractor $x^-_*=\infty$. This case is not interesting,
as the Friedmann constraint demands then that the DM variable $y
\to -\infty$.

The opposite case corresponds to the DE variable located to the
\emph{left} of point C, in which case $x \to -\infty$ and, because
of the Friedmann constraint again, the DM variable grows without
bound $y \to \infty$.

Whether the solution of Eqs.~(\ref{eq:dyna-sysmod2}) is to the
right or to the left of the unstable point C depends on the
initial conditions and the values of the interaction parameters
$\Gamma_I$. For instance, if the universe were described by this
variant of Model~ II, then we would be to the right (left) of
point C if $x_0 > x^+_*$ ($x_0 < x^+_*$).

The special case previously considered in~\cite{Valiviita:2008iv}
corresponds to $\Gamma_x =0$, and then it is a simple realization
of the case just described above. According to our discussion
above, the model needs $\Gamma_c > 0$ in order to have a positive
early DE component, in agreement with~\cite{Valiviita:2008iv}.
(Note that our interaction parameters are defined with an opposite
sign to those of~\cite{Valiviita:2008iv}.)

If $\Gamma_c <0$, the early evolution of the Universe is
problematic, but there exists the late time attractor $x \to 1$
and $y \to 0$, which is precisely the case
in~\cite{Valiviita:2008iv}. However, the case $\Gamma_c >0$ is not
a better option, because point C then becomes an unstable critical
point for which $x_*=1$ and $y_*=0$. According to its present
conditions, our universe would be at the \emph{left} of point C
and its late time evolution would result in $x \to -\infty$ and $y
\to \infty$.

Actually, Eq.~(\ref{eq:motiona}) has an exact solution if
$\Gamma_x = 0$,
\begin{equation}
\rho_c = \rho_{c \, 0}a^{-3} \exp\left[ 3\Gamma_c(t-t_0) \right]
\, , \label{eq:exact1}
\end{equation}
which tells us that the DM energy density is always positivie,
regardless the sign of $\Gamma_c$. It also confirms the
expectations discussed above: (a) if $\Gamma_c <0$, the DM
component may scale at a rate faster than the usual $a^{-3}$ at
early times, but it exponentially vanishes at late times; (b) if
$\Gamma_c >0$, the DM component may scale at a rate slower than
the usual $a^{-3}$ at early times, but it exponentially grows at
late times.

As for the special case $\Gamma_c =0$, we see that the dark
components are well behaved, because both cases $\Gamma_x > 0$ and
$\Gamma_x < 0$ lead the Universe to the (unstable) critical point
$x^+_* = 0$ at early times.

The differences appear at late times. If $\Gamma_x < 0$, the
discussion about the general case $(\gamma_x - \gamma_c) < 0$ also
applies here. For example, our Universe would be presently located
at the right of point C, and then its final fate would be $x \to
\infty$ and $y \to -\infty$.

If $\Gamma_x > 0$, then point C is stable but represents a DM
dominated stage. If this were the case of the present Universe,
this would mean that the present DE dominated epoch is a transient
phenomenon and that the Universe would eventually be dominated by
DM again.

The above statements can be clearly seen from the exact solution
of Eq.~(\ref{eq:motionb}) in the case $\Gamma_c=0$,
\begin{equation}
  \rho_x = \rho_{x\,0}a^{-3(1+w_x)}\exp{[-3\Gamma_x(t-t_0)]} \,
  , \label{eq:exact2}
\end{equation}
which is the companion solution of Eq.~(\ref{eq:exact1}). The
transient character of the DE dominated epoch is controlled by the
interaction $\Gamma_x$; a long enough DE era requires an
appropriately small DE interaction.

\subsection{DE-DM ratio}

The evolution of the DE-DM ratio is governed by
Eq.~(\ref{eq:diffR}), with $\tilde{\alpha}_I \to
  \tilde{\alpha}_I(u)$. We no longer have an exact solution because
of the explicit dependence on $u$; however, there are some
semi-analytical results that can help us to understand the dynamics of
the model.

The troublesome features of negative DE at early times can also be
derived via the equation for $R$:
\begin{equation}
\dot{R} = -3 (w_x H + \Gamma_c + \Gamma_x ) R - 3\Gamma_c -
3\gamma_x R^2 \, . \label{rdot}
\end{equation}
In the matter dominated era, we have $|R| \ll 1$ and
$|\Gamma_I|\ll H$, so that $\rho_c \propto a^{-3}$ and then
$H=2/3t$. Equation~(\ref{rdot}) becomes
\begin{equation}
\dot R \to - \frac{2w_x}{t} R - 3\Gamma_c \, ,
\end{equation}
with solution
\begin{equation}
R \to - \frac{3 \Gamma_c}{1+2w_x} t + C t^{-2w_x} \, ,
\end{equation}
where $C$ is an integration constant that has to be chosen to
impose appropriately the condition $R \to 0$ as $t \to 0$. If
$w_x<-1/2$, the $\Gamma_c$ mode dominates over the $C$ mode as $t
\to 0$, and then the DE becomes negative for $\Gamma_c < 0$.

There is another constraint we may impose on the free parameters
of Model~II. If we require the DE-DM ratio to be a growing
function at the present time, i.e., $R^\prime|_0 > 0$, then
according to Eq.~(\ref{eq:diffR}), we need
\begin{equation}
  \tilde{\gamma}_c \geq - \tilde{\gamma}_x R_0 +
  \frac{R_0}{1+R_0} \, , \label{eq:IIcondition}
\end{equation}
where $\tilde{\alpha}_I(1/2)=\tilde{\gamma}_I$, see
Eq.~(\ref{gamma}).

\begin{figure}[ht]
\includegraphics[width=8.7cm]{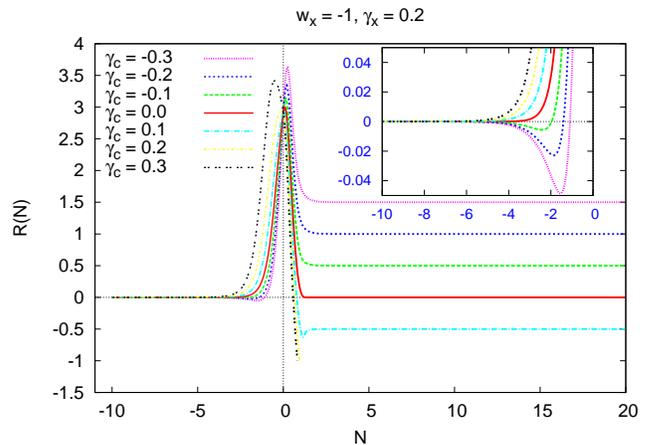}
\caption{\label{fig:model2-1}Examples of the DE-DM ratio $R$ as
  obtained from the numerical solutions of
  Eqs.~(\ref{eq:dyna-sysmod2}), under the condition that all cases
  have $x_0=0.7$ and $y_0=0.24$ at $N=0$. We took a fixed value of
  $\gamma_x = 0.2$ for the case $w_x=-1$, and the values of $\gamma_c$
  are as indicated on the plot. Negative (positive) values of DE at
  early times appear for negative (positive) values of $\gamma_c$ (see
  the inset), and a finite late time attractor appears only if the
  condition $(\gamma_x - \gamma_c)>0$ holds. The values
  for which $(\gamma_x- \gamma_c) \leq 0$ lead to $x \to -\infty$, and
  may also break the constraint~(\ref{eq:IIcondition}).}
\end{figure}

\begin{figure}[ht]
\includegraphics[width=8.7cm]{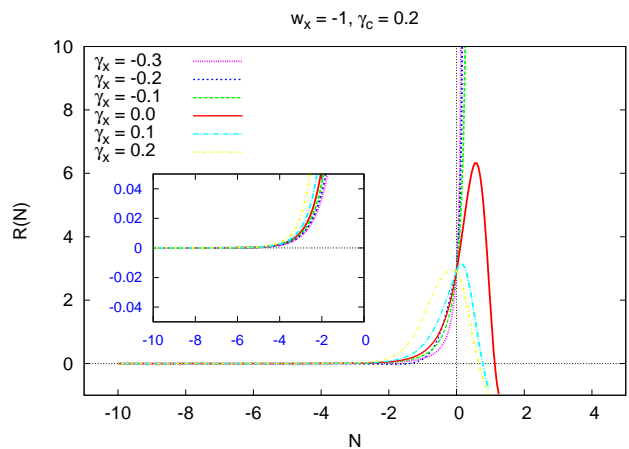}
\caption{\label{fig:model2-2} The same as in
Fig.~\ref{fig:model2-1},
  but now for the fixed value $\gamma_c=0.2$. As expected, early
  evolution is positive because $\gamma_c > 0$ (see the inset), but
  point C is unstable in all cases. The DE-DM ratio first diverges for
  $x_0 < x^+_*$, but asymptotes to $R \to \pm -1$ at late times
  whatever the case $x_0 > x^+_*$ or $x_0 < x^+_*$, see
  Sec.~\ref{sec:special-cases-} and Eqs.~(\ref{eq:III-asymptotic1b}).}
\end{figure}

We show in Figs.~\ref{fig:model2-1} and~\ref{fig:model2-2} some
numerical examples for the evolution of the DE-DM ratio obtained
from Eqs.~(\ref{eq:dyna-sysmod2}). The different cases confirm the
results on Model~II as discussed above.

In particular, if we require positivity of both dark compoinents
as $t\to\infty$ and as $t\to0$, then we require $\Gamma_c = 0$ and
a positive small value of $\Gamma_x$; the smaller $\Gamma_x$ is,
the larger the maximum value reached by the DE-DM ratio before the
Universe enters again into a late-time DM dominated epoch.

\section{Conclusions \label{sec:conclusions}}

We have made a careful analysis of two simple models of
interaction between DM and DE. The first model, Model~I, has an
interaction term proportional to the Hubble parameter times a
linear combination of the dark sector densities, and is one of the
most studied in the literature.

We developed different mathematical techniques to investigate the
properties of the model under simple but general enough
assumptions. Our results recover those of previous studies, and we
found new analytic expressions that clarify the limitations of
Model~I.

To begin with, we absorbed the (constant) DE equation of state
$w_x$ into the equations of motion, so that the parameter-space is
truly 2-dimensional; this very much simplified the study of the
allowed values of the interaction parameters (compare our
Fig.~\ref{fig:parameters} with the corresponding 3-dimensional
figures in~\cite{Quartin:2008px}).

The absorption of the equation of state is not a mere mathematical
trick. The original equations of motion would have been
\begin{subequations}
  \begin{eqnarray}
    x^\prime &=& -3 w_x x \left( 1 - x \right) - z \, , \\
    y^\prime &=& 3w_x xy + z \, .
\end{eqnarray}
\end{subequations}
We can see that Eqs.~(\ref{eq:dyna-sys}) are recovered by setting
$w_x=-1$ in the above equations. Thus, the cases discussed are in some
sense isomorphic to the cosmological constant interacting case.

This fact also shows the degeneracy between interacting models
with a constant DE equation of state; if there is a successful
model with a cosmological constant, one can find another one with
a different value of $w_x$. However, the models are
distinguishable in their particular evolution, as the DE equation
of state should appear explicitly in the final expression of the
DE-DM ratio, see Eq.~(\ref{eq:simple-ratio}), and other
quantities.

Even though our general assumption was $w_x < 0$, it is clear
  that in practice we have to restrict ourselves to values of the DE
  equation of state that allow an accelerating expansion at the
  present time, i.e. $w_{\rm tot} < -1/3$. It is also possible to
  allow \emph{phantom} values $w_x<-1$, although it is not clear
  whether any physically consistent models exist in this regime. A
  comparison with observations would then be required to distinguish
  among the different possibilities, but this is beyond the purposes
  of the present study.

We presented for the first time the surprisingly simple
equation~(\ref{DEtoDM}) for the DE-DM ratio, and its simple
analytic solution~(\ref{eq:simple-ratio}). We stress that the
exact solution~(\ref{eq:simple-ratio}) allows us to easily and
directly uncover the limitations of this interaction model. The
exact solution has only one free integration constant, and thus we
need a careful choice of initial conditions ensure an evolution
that remains close to that of the standard non-interacting
cosmology.

Surprisingly restrictive limitations are needed on the interaction
parameters, $\alpha_c$ and $\alpha_x$. Our simplified study of the
parameter space showed that the parameters should take very small
values. This is also in agreement with other studies in which one
can find a careful comparison of the model with cosmological
observations (see for
instance~\cite{Quartin:2008px,delCampo:2008sr,delCampo:2008jx,Olivares:2007rt,Guo:2007zk}). We
showed that the model that seems to work better is the simple
case $\alpha_c =0$. This model is almost indistinguishable from
the standard $\Lambda$CDM at early times, but provides a finite
DE-DM ratio at late times so that the coincidence problem is
alleviated.

Other interesting properties arise from the interacting Model~II
of Sec.~\ref{sec:model-ii.-q}. Even though it can be thought of as
a time-dependent version of Model I, the properties of the
critical points differ significantly.

First of all, its vanishing interaction variables at early times
allow Model~II to have a true matter dominated epoch, which in
fact corresponds to a (unstable) critical point of the dynamical
system. However, a positive DM interaction $\Gamma_c > 0$ is
necessary in order to keep the DE component positive at early
times.

Also, the DE-DM ratio can be finite and positive at late times,
but this requires the dark interactions to have opposite signs and
to comply with the condition $\Gamma_x > \Gamma_c$. It is then
apparent that we cannot, in general, have a realization of
Model~II in which the dark components are well behaved for all
times and the coincidence problem is addressed.

There is though a simple model that may be realistic and
corresponds to $\Gamma_c =0$. This model can satisfy all the
constraints and the DE interaction $\Gamma_x$ can be adjusted so
that the Universe can have appropriate matter and DE eras. The
only difference is that DE domination is a transient event: the
Universe would eventually go back to a DM dominated epoch at late
times.

Finally, we note that the problem of negative dark sector densities is 
not the only problem with Models~I and II: the curvature perturbation 
also has a non-adiabatic instability on large 
scales~\cite{Valiviita:2008iv}. As pointed out 
in~\cite{Valiviita:2008iv}, both of these pathologies in the 
interaction models are related to the way in which we treat DE, i.e. as 
a fluid with constant $w_x$. If $w_x$ is allowed to vary in the early 
universe (as happens with scalar field DE), then the pathologies can be
avoided.

\begin{acknowledgments}
GC-C is supported by the Programme Alban (the European Union
Programme of High Level Scholarships for Latin America),
scholarship No. E06D103604MX, and the Mexican National Council for
Science and Technology, CONACYT, scholarship No. 192680. The work
of RM was supported by the UK's Science \& Technology Facilities
Council, and by a Royal Society exchange grant. The work of LAU-L
is supported by CONACYT (56946), DINPO and PROMEP-UGTO-CA-3.
RM thanks the Departmento de F\'isica at the Universidad de Guanajuato
for hospitality during part of this work. LAU-L would like to thank
Diana Ju\'arez and Daniel de la Torre for very helpful discussions.
\end{acknowledgments}

\bibliography{dm-derefs}

\end{document}